\documentstyle[12pt,epsfig]{article}
\begin{document}
\vspace{3cm}
\begin{center}

  {\bf   SPIN EFFECTS AND RELATIVE MOMENTUM
         SPECTRUM OF TWO PROTONS IN DEUTERON
 CHARGE-EXCHANGE BREAKUP  }\\

\vspace{0.7cm}

{\bf R.Lednicky$^a$, V.L.Lyuboshitz$^b$, V.V.Lyuboshitz$^b$}\\

\vspace{0.7cm}

  {\it Joint Institute for Nuclear Research\\
  141980, Dubna, Moscow Region, Russia\\
  E-mail:\\ ${}^a$lednicky@fzu.cz,
 ${}^b$lyubosh@sunhe.jinr.ru    }\\

\vspace{0.7cm}
\end{center}
 \begin{abstract}
       The relation between
   the differential cross section of the charge-exchange breakup of a fast
   deuteron $ d + p \rightarrow (pp) + n$ and the differential cross section
   of the charge transfer process $ n + p \rightarrow p + n $
   is discussed taking into account
   the effects of the proton identity (Fermi-statistics) and
   of the Coulomb and strong interactions of protons in the final state.
   The distribution of the relative momenta of the protons produced
   in the charge-exchange process $ d + p \rightarrow (pp) + n $ in the
   forward direction is found in the framework of the impulse approach.
   At the momentum transfer close to zero, the differential cross section of
   the charge-exchange breakup of a fast
   deuteron on a proton target is determined only by
   the spin-flip part of the amplitude of the charge transfer reaction
    $ n + p \rightarrow p + n $ at zero angle. The dependence of the
   differential cross section of the process $ d + p \rightarrow (pp)
   + n $ in the forward direction on the polarization parameters of
   the deuteron and the proton is investigated. It is shown that the
   study of the process $ d + p \rightarrow (pp) + n $ with a beam of
   polarized (aligned) deuterons on a polarized proton target
   allows one, in principle, to separate
   two spin-dependent terms in the amplitude of the charge transfer reaction
   $ n + p \rightarrow p + n$ at zero angle and
   to determine their phase difference. The influence
   of the deuteron $d$-wave state on the polarization effects
   and the spectrum of the relative momenta of two protons
   produced in forward direction in the charge-exchange
   breakup of a polarized deuteron on a polarized proton target
   is investigated.\\
   {\bf Key-words}: deuteron, proton, charge-exchange breakup,
    spin, polarization, momentum spectrum.
\end{abstract}

\section {Introduction}

 Earlier [1,2] the charge-exchange breakup of a fast
   deuteron into two protons
$$
 d + a \rightarrow (pp) + b
 $$
 was investigated in the framework of the
impulse approach. In this process the neutron incorporated in
the deuteron is converted into the proton as a result of the charge
 transfer reaction
 $$
 n + a \rightarrow p + b
 $$
 at the momentum $ {\bf p}_n = {\bf p}_d/2$. In so doing,
 it was supposed that the
 target particles $a$ and $b$ with the unit charge difference are
the members of the same isomultiplet. In particular, one can
consider the process $$ d + p \rightarrow (pp) + n $$
taking place at a collision of a fast deuteron with a proton
target.

In papers [1,2] only the case of unpolarized initial particles was
considered. In so doing, the interaction between the final protons was
not taken into account and thus the spectrum of proton
relative momenta was determined incorrectly.
In paper [3], a more detailed analysis of the problem
has been carried out, taking into account
the polarization parameters of the particles,
the identity of the produced protons (Fermi-statistics effect) and the
Coulomb and strong interactions in the final state.
The latter was however treated in an approximate way with the
accuracy of some tens percent.
Here we calculate this effect on a percent level based
on the exact solution of the Schroedinger equation.
We also calculate the influence of the d-wave state in the
deuteron wave function on the proton spectrum and
polarization effects
in the forward charge-exchange deuteron breakup.

\section {Impulse approach}

The impulse approach to the deuteron breakup is applicable, in any case,
for relativistic energies guaranteeing validity of the condition [3]
\begin{equation}
 v \gg  \sqrt{\epsilon/m} \sim 1/20,
\end{equation}
where $v$ is the velocity of the projectile
deuteron, $\epsilon$ is the binding energy of the nucleons in the
deuteron, $m$ is the nucleon mass. The duration of the collision is then
much smaller than the characteristic period of the motion of the nucleons
in the deuteron and, as a result, the neutron and the proton have
practically no time to change their coordinates during the impact.

When the neutron, at the transition into the proton, gets
the nonrelativistic momentum transfer {\bf q}
in the rest frame of the deuteron,
then the wave function of the two protons immediately after the impact will
have the form (we use the unit system with $\hbar = c = 1$):
\begin{equation}
    \psi_{{\rm pp}} = \exp(i{\bf qR})\, \psi_{{\rm d}}({\bf r})
\, \exp(-i{\bf qr}/2),
\end{equation}
where ${\bf r} = {\bf r}_{{\rm p}} - {\bf r}_{{\rm n}}$ is the coordinate
of the relative motion, ${\bf R}  = ({\bf r}_{{\rm p}}
+ {\bf r}_{{\rm n}})/2$ is the coordinate of the center-of-mass,
$\psi_{{\rm d}}({\bf r})$ is the deuteron wave function.

The expansion of the function
\begin{equation}
\psi({\bf r}) = \psi_{{\rm d}}({\bf r})\, \exp(-i{\bf qr}/2)
\end{equation}
over the stationary wave functions of the two protons, taking into
account their identity and interaction in final state,
gives the continuous spectrum
of the relative momenta of the created protons. The magnitude of
the differential cross section of the charge-exchange breakup of the
deuteron is determined by the transitions from the deuteron spin
states to the spin states of the two-proton system. In
the impulse approximation, the contributions of these
transitions are connected with the spin structure of the amplitude
of the charge transfer reaction $ n + a \rightarrow p + b $.

\section {Spin structure of the nucleon charge transfer\\ reaction}

 The amplitude of the process $n + a \rightarrow p + b$ has the
structure
\begin{equation}
  \hat{f}( n + a \rightarrow p + b ) = [\,\hat {C}(t) +
\hat{{\bf B}}(t) \hat{\mbox{\boldmath $\sigma$}}\,] P_{{\rm exch}},
\end{equation}
where $t$ is the four-momentum transfer squared;
$P_{\rm exch}$ is the exchange operator transforming the neutron
into the proton and the target particle $a$ into the particle $b$;
$\hat{\mbox{\boldmath $\sigma$}}$ is the Pauli vector operator,
 acting on the
spin states of the neutron and the proton; $\hat{C}(t)$ and
$\hat {{\bf B}}(t)$ are the operators acting on the spin states
of the target particles $a$ and $b$.

The differential cross section of the reaction $n + a \rightarrow
p + b$ in the case of unpolarized particles has the form
\begin{equation}
\frac{d\sigma}{dt}(n + a \rightarrow p + b) =
\frac{d\sigma}{dt}^{({\rm nf})}( n + a \rightarrow p + b) +
\frac{d\sigma}{dt}^{({\rm f})}( n + a \rightarrow p + b),
\end{equation}
where
\begin{equation}
  \frac{d\sigma}{dt}^{({\rm nf})}(n + a \rightarrow p + b)
 =\frac{1}{2j + 1}{\rm tr}\, [\hat{C}(t) \hat{C}(t)]
\end{equation}
is the {\it spin-nonflip} part of the cross section, which is not
connected with the spin quantum numbers of the fast nucleons, and
\begin{equation}
\frac{d\sigma}{dt}^{({\rm f})}(n +a \rightarrow p + b)
= \frac{1}{2j + 1} {\rm tr}\, [\hat{{\bf B}}(t)\hat{{\bf B}}^+(t)]
\end{equation}
is the {\it spin-flip} (spin-dependent) part of the cross section,
conditioned by the presence of the nucleon spin. Here the symbol
tr ({\it trace}) denotes the sum of the diagonal elements of the
operators acting in the spin space of target particles $a$
and $b$; $j$ is the spin of the target.

\section {Charge-exchange breakup of the unpolarized \\
deuteron on an unpolarized target}

 As it is known, the neutron and the proton in the deuteron
are in the triplet state (with the unit total spin). When the
deuteron is unpolarized, then each of the three spin states,
corresponding to the projections of the total spin onto the
quantization axis $z$ equal to +1,--1 and 0, is realized
with the probability of 1/3. These states are the following:
 $$
 \mid \chi^{(t)}_{+1}\rangle = \mid +1/2\rangle^{(1)}
 \cdot \mid + 1/2 \rangle^{(2)} ;\quad
  \mid \chi^{(t)}_{-1}\rangle = \mid -1/2 \rangle^{(1)}
 \cdot \mid -1/2 \rangle^{(2)} ;
 $$
\begin{equation}
   \mid \chi^{(t)}_{0}\rangle = \frac{1}{\sqrt{2}} \left(\mid
+1/2\rangle^{(1)} \cdot \mid - 1/2 \rangle^{(2)} +
 \mid -1/2 \rangle^{(1)} \cdot \mid +1/2 \rangle^{(2)}\right)  .
\end{equation}
Here the index 1 is related to the spin function of the neutron,
and the index 2 is related to the spin function of the spectator
proton. In the process $ d + a \rightarrow (pp) + b $ the system
of two protons can be created in the triplet states (the index 1
in Eqs. (8) then also corresponds to a proton) as well as
in the singlet state with the zero total spin:
\begin{equation}
   \mid \chi^{(s)}_{0}\rangle = \frac{1}{\sqrt{2}} \left(\mid
+1/2\rangle^{(1)} \cdot \mid - 1/2 \rangle^{(2)} -
 \mid -1/2 \rangle^{(1)} \cdot \mid +1/2 \rangle^{(2)}\right)  .
\end{equation}
It is easy to see that the operator $\hat{C}(t)$ in Eq. (5), being
independent of the nucleon spin, leads to the production of the
$(pp)$-system only in the triplet states. As for the
{\it spin-flip}
operator $\hat{{\bf B}} \hat{\mbox{\boldmath $\sigma$}}$, it
generates the transitions to both the triplet and singlet states.

In accordance with the Pauli principle for identical fermions, the
coordinate wave function of two protons in the singlet (triplet)
state is symmetric (antisymmetric) and the orbital angular
momenta have only even (odd) values [4]. Taking
into account the proton identity and the properties of the Pauli
matrices, the differential cross section of the process
$d + a \rightarrow (pp) + b$ for unpolarized particles is expressed
through the {\it spin-nonflip}  and {\it spin-flip} parts of the
 differential cross section of the reaction $n + a \rightarrow p + b$
 at the momentum ${\bf p}_n = \frac12 {\bf p}_d$ [3]:
$$
\frac{d^4\sigma}{dt} (d +a \rightarrow (pp) + b) =
\Bigl\{\Bigl[\frac{d\sigma}{dt}^{(nf)}(n + a \rightarrow p +b) +
\frac{2}{3}\,\frac {d\sigma}{dt}^{(f)}(n + a \rightarrow p + b)
\Bigr]\times
 $$
 $$ \times \Bigl| \int\psi_d({\bf r})\exp(-i{\bf qr}/2)
\,\frac{1}{\sqrt{2}}\,\left(\phi^{*(-)}_{{\bf k}}({\bf r}) -
\phi^{*(-)}_{\bf k}(-{\bf r} \right) d^3 {\bf r}\,\Bigr|^2 +
$$
\begin{equation}
 + \frac{1}{3}\,\frac {d\sigma}{dt}^{(f)}(n + a \rightarrow p + b)
  \, \Bigl| \int\psi_d({\bf r})\exp(-i{\bf qr}/2)
\,\frac{1}{\sqrt{2}}\,\left(\phi^{*(-)}_{{\bf k}}({\bf r}) +
\phi^{*(-)}_{\bf k}(-{\bf r} \right) d^3 {\bf r}\, \Bigr|^2
 \Bigr\}\frac{d^3 {\bf k}}{(2\pi)^3}.
\end{equation}
Here $-{\bf k}$ is the three-momentum of the spectator proton in the
c.m.  frame of the proton pair (it is assumed that the respective
momenta $\pm {\bf k}+\frac12 {\bf q}$ of the active and spectator
protons in the deuteron rest frame  are nonrelativistic, i.e.  $k
 = \mid {\bf k} \mid \ll m,\, \mid {\bf q} \mid
\ll 2 m$); $\phi^{(-)}_{{\bf k}}({\bf r})$ is the nonsymmetrized wave
function of the relative motion of the two interacting protons
corresponding to the scattering problem and having the asymptotics
in the form of the superposition of the plane wave and the converging
spherical wave. Let us emphasize that in Eq. (10), the
antisymmetrization and the symmetrization of the wave function
$\phi^{(-)}_{{\bf k}}({\bf r})$ with respect to the substitution
${\bf r}\rightarrow -{\bf r}$ are performed for the transitions
to the triplet states and to the singlet state, respectively.

In the case of forward production of the two-proton system,
we can put the value $t \approx -{\bf q}^2 =0$.\footnote
{$^)$ The momentum transfer due to
the change of the effective mass of the two nucleons at the transition
$ d \rightarrow pp $ is negligibly small as compared with the inverse
radius of the deuteron.}$^)$ As a result, the contribution of the
transitions to the two-proton triplet states vanishes
since the integrated function in the first term in Eq. (10) then
represents the product of the antisymmetric two-proton
coordinate wave function and the symmetric deuteron one.
Thus, the differential cross section of the forward charge-exchange
breakup of the unpolarized deuteron on an unpolarized target
is proportional to the {\it spin-flip}  part of the differential
cross section of the charge transfer reaction $ n + a \rightarrow
 p + b $ at zero angle [3]:
\begin{equation}
\frac{d^4 \sigma}{dt} (d + a \rightarrow (pp) + b)\Bigl |_{t=0} =
\frac{2}{3}\, \frac{d\sigma}{dt}^{(f)}(n + a \rightarrow p + b)
\Bigl |_{t=0}\, \cdot \, \Bigl| \int \psi_d({\bf r})\, \phi^{*(-)}_{{\bf
k}} ({\bf r})\,d^3 {\bf r} \Bigr|^2 \frac{d^3{\bf k}}{(2\pi)^3}.
\end{equation}

Generally, for any two-proton production angle, one can
use the completeness conditions for the two-proton
wave functions in continuous spectrum:
\begin{equation}
  \frac{1}{(2\pi)^3} \int \phi^{*(-)}_{{\bf k}}({\bf r})
\,\phi^{(-)}_{{\bf k}}(\mp {\bf r'})\, d^3 {\bf k} =
\delta^3({\bf r} \pm {\bf r'})
\end{equation}
and integrate the differential
cross section over the proton three-momentum ${\bf k}$
in the two-proton rest frame:
\begin{equation}
\frac{d\sigma}{dt} (d + a \rightarrow (pp)  + b) =
\frac{d\sigma^{(nf)}}{dt}(n + a \rightarrow p + b)
\Bigl( 1 - F({\bf q})\Bigr)+
\frac{d\sigma^{(f)}}{dt}(n + a \rightarrow p + b)
\Bigl( 1 -\frac{1}{3} F({\bf q})\Bigr),
\end{equation}
where
$$
  F({\bf q}) = \int \Bigl(\psi_d ({\bf r})\Bigr)^2 \,\exp (-i{\bf
qr})\, d^3{\bf r}
$$
is the deuteron form-factor.
Particularly, in forward direction
($ t\approx -{\bf q}^2 = 0,\, F(0) = 1$)
the impulse approximation in Eq. (13)
 leads to the simple relation [3]:
\begin{equation}
\frac{d\sigma}{dt}(d + a \rightarrow (pp) + b) \Bigl|_{t=0}=
\frac{2}{3}\,\frac{d\sigma^{(f)}}{dt}(n + a \rightarrow p + b)
\Bigl|_{t=0}.
\end{equation}

In previous formulae we have neglected the contribution of the
deuteron $d$-state, considering only the $s$-wave part
of the deuteron wave function
(in this case $\psi_d ({\bf r}) = \psi_d (r)$, where
$ r = |{\bf r}|$). However, we will show later on that Eq. (14)
for the cross section integrated over the spectrum of the proton
relative momenta is valid also in the general case
when the deuteron $d$-state
is taken into account.

\section {Forward charge-exchange breakup of the polarized deuteron
on the polarized proton target}

Following paper [3], we will account now for the possible
projectile and target polarization states in the process
$ d + p \rightarrow (pp) + n$  in forward direction.
The amplitude of the charge transfer reaction $n +p \rightarrow
p + n$ at zero angle can be represented in the following
general form:
\begin{equation}
\hat {f} = \hat {P}_{\rm {exch}}\Bigl[ c + b \Bigl(
\hat {\mbox {\boldmath $\sigma$}}\hat {\mbox {\boldmath $\sigma$}^{(1)}}
-(\hat {\mbox {\boldmath $\sigma$}}{\bf l})(\hat {\mbox {\boldmath
$\sigma$}^{(1)}}{\bf l})\Bigr)
+ a\,(\hat {\mbox {\boldmath $\sigma$}}{\bf l})(\hat {\mbox {\boldmath
$\sigma$}^{(1)}}{\bf l})\Bigr)\Bigr],
\end{equation}
where $\hat {\mbox {\boldmath $\sigma$}^{(1)}}$ is the vector Pauli
operator acting on the spin states of the target nucleons,
${\bf l}$  is the unit vector parallel to the neutron momentum.
In the
considered case the operator $\hat{{\bf B}}(0)$ in the general
Eq. (4) has the form
\begin{equation}
\hat{{\bf B}}(0) = b\, \hat {\mbox {\boldmath $\sigma$}^{(1)}} +
(a - b)\, {\bf l}\, (\hat {\mbox {\boldmath
$\sigma$}^{(1)}}{\bf l}).
\end{equation}
In accordance with Eqs. (7), (11), (16), taking into account
only the $s$-wave state of the deuteron,
the differential cross section
of the forward charge-exchange breakup of the unpolarized fast deuteron
on the unpolarized proton (hydrogen) target takes the form:
\begin{equation}
 \frac{d^4\sigma }{dt}(d + p \rightarrow (pp) + n)\Bigl|_{t=0} =
\frac{2}{3}\, \left( 2|b|^2 + |a|^2 \right)\,
\Bigl|\int \psi_d(r)\, \phi^{*(-)}_{{\bf k}} ({\bf r})\, d^3 {\bf r}
\Bigr|^2 \, \frac{d^3 {\bf k}}{(2\pi)^3}.
\end{equation}
The factor in brackets represents the explicit form of the
{\it spin-flip} part of the differential cross section of the
charge transfer reaction $n +p \rightarrow p + n$ at zero angle.
The account of the deuteron and target proton polarizations
yields [3]:
\begin{equation}
 \frac{d^4\sigma }{dt}(d + p \rightarrow (pp) + n)\Bigl|_{t=0} =
 \frac {d\sigma}{dt}(d + p \rightarrow (pp) + n)
\Bigl|_{t=0}\,G({\bf k})\,\frac{d^3 {\bf k}}{(2\pi)^3},
\end{equation}
where
  $$
\frac{d\sigma}{dt}( d + p \rightarrow (pp) + p) \Bigl|_{t=0} =
 2\,\Bigl[\frac{2|b|^2 + |a|^2}{3} +
t^{(d)}_{2\,0}\, ( |b|^2 - |a|^2 ) - |b|^2 (P^{(d)}_{\parallel}
P^{(p)}_{\parallel})-
$$
\begin{equation}
 - {\mathrm Re}(ba^*)\,({\bf P}^{(d)}_{\perp}{\bf P}^{(p)}_{\perp})
+ {\mathrm Im} (ba^*)\, ( {\bf t}_{\perp}^{(d)} [ {\bf P}^{(p)} {\bf l} ])\Bigr],
\end{equation}
\begin{equation}
G({\bf k})=
\Bigl|\int \psi_d(r)\, \phi^{*(-)}_{{\bf k}} ({\bf r})\, d^3 {\bf r}
\Bigr|^2 .
\end{equation}
Here $P^{(p)}_{\parallel} = {\bf P}^{(p)}{\bf l}$\, and\,
$ {\bf P}^{(p)}_{\perp} = {\bf P}^{(p)} - {\bf l}({\bf P}^{(p)}{\bf l})$
are the longitudinal and transverse components of the polarization
vector of the proton, $P^{(d)}_{\parallel}$\, and ${\bf
P}^{(d)}_{\perp}$ are the analogous parameters of the deuteron
vector polarization; ${\bf t}_{\perp}^{(d)}={\bf t}^{(d)}-
{\bf l}({\bf t}^{(d)}{\bf l})$,
${\bf t}^{(d)}$ is the average value of the
operator
 \begin{equation} \hat{{\bf t}} = \hat{{\bf s}}(\hat{{\bf
s}}{\bf l}) + (\hat{{\bf s}}{\bf l}) \hat{\bf s},
 \end{equation}
where $\hat{\bf s} = \{\hat{s}_x, \hat{s}_y, \hat{s}_z \}$ is the operator
of the deuteron spin,  $t_{2\,0} =({\bf t}{\bf l})/2 - 2/3$.
It should be stressed that the operator $\hat{{\bf t}}$ corresponds to
the tensor polarization of the deuteron.

In the coordinate system $\{x, y, z\}$ with the quantization axis
$z$ parallel to
the direction ${\bf l}$ of the deuteron momentum, the deuteron polarization
parameters are expressed through the elements of the spin density
matrix $\rho^{(d)}_{\mu,\nu}$ in the following form:
$$
  P^{(d)}_{\parallel} \equiv P^{(d)}_z = \rho^{(d)}_{+1,+1}
 - \rho^{(d)}_{-1,-1},\,
P^{(d)}_x = \sqrt{2}\,{\mathrm Re} \left(\rho^{(d)}_{+1,0} +
 \rho^{(d)}_{-1,0}\right), \,
P^{(d)}_y = - \sqrt{2}\,{\mathrm Im} \left(\rho^{(d)}_{+1,0} -
 \rho^{(d)}_{-1,0}\right);
$$
$$
t_{2\,0} = \frac{1}{3} - \rho^{(d)}_{0,0},\quad t^{(d)}_z =
2\left(1 - \rho^{(d)}_{0,0}\right),
$$
\begin{equation}
t^{(d)}_x = \sqrt{2}\,{\mathrm Re} \left(\rho^{(d)}_{+1,0} -
 \rho^{(d)}_{-1,0}\right), \quad
t^{(d)}_y = - \sqrt{2}\,{\mathrm Im} \left(\rho^{(d)}_{+1,0} +
 \rho^{(d)}_{-1,0}\right).
\end{equation}
Particularly, Eqs. (18)-(20) yield at $a = b$:
\begin{equation}
\frac{d^4\sigma }{dt}(d + p \rightarrow (pp) + n) \Bigl |_{t=0} =
2|b|^2 \left( 1 - ({\bf P}^{(p)}{\bf P}^{(d)})\right)
G({\bf k})\,\frac{d^3{\bf k}}{(2\pi)^3}
\end{equation}
and, in the case of the unpolarized proton target:
\begin{equation}
\frac{d^4\sigma}{dt} (d + p \rightarrow (pp) + n)\Bigl |_{t=0} =
2\,\left[\frac{2|b|^2 + |a|^2}{3} + t^{(d)}_{2\,0} (|b|^2 -
|a|^2) \right]\, G({\bf k})\,\frac{d^3{\bf k}}{(2\pi)^3}.
\end{equation}
Thus the forward charge-exchange deuteron breakup is sensitive
to the longitudinal tensor polarization of the deuteron if only
$a \ne b $.

We may conclude that the study of the forward breakup process
$d + p \rightarrow (pp) + n$ in
a beam of polarized (aligned) deuterons on the unpolarized hydrogen
target allows one, in principle, to separate the two spin-dependent
terms $a$ and $b$ in the amplitude of the charge transfer reaction
$ n + p \rightarrow p + n$ at zero angle, one of them ($b$)
not conserving the projection
of the nucleon spin onto the momentum direction at the transition
of the neutron into the proton. When both the deuteron and the
proton are transversely polarized, it is possible to determine the
phase difference between the amplitudes $b$ and $a$.

\section {Spectrum of proton momenta in the forward
charge-exchange deuteron breakup}

In accordance with Eq. (18), when taking into account only
the $s$-wave part of the deuteron wave
function, the proton momentum spectrum in the forward
charge-exchange deuteron breakup is factorized in the function
$G({\bf k})$, defined in Eq. (20), independently of
the polarization parameters of the deuteron and the proton.
The function $G({\bf k})$ determines the proton spectrum in the
two-proton rest frame which, in the considered case of the forward
deuteron breakup, coincides with the deuteron rest frame.
This function satisfies the completeness condition
 \begin{equation} \int G({\bf k})
\frac{d^3{\bf k}}{(2\pi)^3} = \int\, \Bigl|\int \psi_d({\bf r})\,
\phi^{*(-)}_{{\bf k}} ({\bf r})\, d^3 {\bf r} \Bigr|^2 \, \frac{d^3
{\bf k}}{(2\pi)^3} = 1.
  \end{equation}
Since here we consider only the $s$-wave state of the deuteron,
we can write $\psi_d({\bf r}) = \psi_d(r),\\ G({\bf k}) = G(k)$,
where $ r=|{\bf r}|,\, k = |{\bf k}|$.

In paper [3], the analytical expression for the function $G(k)$
was obtained based on the approximate formula for the two-proton
wave function at the distances $r\ll a_B= 57.5$ fm, $ r<\sim 1/k$.
This expression violates the completeness condition (25) and its
accuracy is on the level of tens percent.
Here we will calculate the function $G(k)$ based on the exact
solution of the Schroedinger equation describing correctly
the parameters of the low-energy $pp$-scattering.

Let us represent the normalized wave function of the deuteron
$s$-wave state in the form
\begin{equation}
 \psi_d(r) = \frac{1}{\sqrt{4\pi}}\,
\frac{u(r)}{r},~~~
\int\limits_0^{\infty}\bigl(u(r)\bigr)^2\, dr = 1
 \end{equation}
and expand the two-proton wave function over
the partial waves [4]:
\begin{equation}
\phi_{{\bf k}}^{(-)}({\bf r}) = 4\pi \sum_l i^l \exp\bigl(-i\delta_l(k) \bigr)
\frac{R_{kl}(r)}{kr}\sum_m Y^*_{lm}\left(\frac{{\bf k}}{k}\right)
Y_{lm}\left(\frac{{\bf r}}{r}\right),
\end{equation}
where $\delta_l(k)$ is the partial phase of the elastic scattering,
$Y_{lm}({\bf k}/k)$ and $Y_{lm}({\bf r}/r)$ are the spherical functions,
$R_{kl}(r)$ are the real radial functions satisfying the completeness
condition
$$
\int\limits_0^{\infty} R_{kl}(r)\, R_{kl}(r') \,dk = \frac{\pi}{2}
\delta(r - r').
$$
Substituting Eqs. (26) and (27) into Eq. (20), we get
\begin{equation}
  G(k) = \frac{4\pi}{k^2} \Bigl( \int \limits_0^{\infty}
u(r)\,R_{k0}(r)\,dr\Bigr)^2.
\end{equation}
Thus, only the $s$-wave part of the two-proton wave function
contributes into the factor $G(k)$.
Taking further into account the isotropy of the momentum distribution,
we can rewrite Eq.~(18) in the form
\begin{equation}
 \frac{d^2\sigma }{dt}(d + p \rightarrow (pp) + n)\Bigl|_{t=0} =
 \frac {d\sigma}{dt}(d +p \rightarrow (pp) + n) \Bigl|_{t=0}
 \Bigr( \int \limits_0^{\infty}
u(r)\,R_{k0}(r)\,dr\Bigr)^2 \frac{2}{\pi}\,dk.
\end{equation}
In the numerical calculations, we will use
for the normalized deuteron $s$-wave function $u(r)$
the Hulthen expression as well as
the analytical solution corresponding
to the simple square-well potential.
The strong square-well potential will be also used to calculate
the two-proton $s$-wave radial function $R_{k0}(r)$.

The normalized Hulthen wave function with the correct asymptotic
behavior has the form [5]
\begin{equation}
u(r)=  \sqrt{\frac{2}{\rho - d}}\left(e^{-r/\rho} - e^{-\alpha
r/\rho}\right),
\end{equation}
  where $\rho = 1/\sqrt{m\epsilon}$ = 4.3 fm is the deuteron
radius, $d$ = 1.7 fm is the effective radius, $ \alpha \approx 6.25$
(see, for example, [6]).

In the case of the square-well potential $-V$ of a width $b$,
corresponding to the
deuteron binding energy $\epsilon$ = 2.3 MeV
of the triplet $s$-wave proton-neutron state, we have
\begin{equation}
u(r) = \sqrt{\frac{2}{\rho - d}}\, e^{-r/\rho},\, r>b; \qquad
   u(r)= \sqrt{\frac{2}{\rho - d}}\, e^{-b/\rho}\, \frac{\sin \kappa r}
 {\sin\kappa b}, \, r<b .
\end{equation}
    Here $b$ = 2.06 fm, $\kappa =
\sqrt{m(V - \epsilon)}$ = 0.883 fm$^{-1}$, $V$ = 34 MeV.

The $s$-wave radial function of two protons in singlet state
in the potential being the
sum of the Coulomb potential and the strong square-well potential
$-V$ of a width $b$ is the following [7]:
  $$
    R_{k0}(r) = F_0(k,r)
  \cos \tilde{\delta}_0 (k) +
  G_0(k,r)\sin \tilde{\delta}_0 (k),\,r>b;
    $$
    $$
    R_{k0}(r) =
 \frac{F_0(\tilde{\kappa},r)}{F_0(\tilde{\kappa},b)} \left[F_0(k,b)
 \cos \tilde{\delta}_0 (k) +
 G_0(k,b)\sin \tilde{\delta}_0 (k)\right],\,r<b.
    $$
   Here $\tilde{\delta}_0$ is the $s$-wave phase connected with the
   short-range potential, $\tilde{\kappa} =\sqrt{mV + k^2}$,
 $F_0$ and $G_0$ are the regular and irregular Coulomb functions,
   respectively [4]. In so doing,
   \begin{equation}
  \cot \tilde{\delta}_0(k) \doteq
   \frac{1}{kA_c}\left( \frac{1}{f_0} + \frac{1}{2} d_0k^2 -
\frac{2}{a_B} h(ka_B)\right),
\end{equation}
   where
   \begin{equation}
      A_c = \frac{2\pi}{ka_B}
      \left[\exp\left(\frac{2\pi}{ka_B}\right) - 1\right]^{-1}
 \end{equation}
 is the Coulomb (Gamov) factor describing the Coulomb repulsion of
   the protons at zero separation,
   $a_B = (e^2m/2)^{-1}$ =57.5 fm is the Bohr radius
   of the two-proton system, $f_0$ is the scattering length, $d_0$
   is the effective radius; the function $h(y)$ is given by the
    formula
 $$
h(y) = \sum \limits_{n =  1}^{\infty} \frac{1}{n(n^2y^2 +
   1)} -  C + \ln y
 $$
    ($C$ = 0.577... is the Euler constant). The
 following values of the parameters correspond to the experimental
 data on the low-energy proton-proton scattering: $V$ = 11.7 MeV, $b$
= 2.78 fm, $f_0$ = 7.8 fm, $d_0$ = 2.8 fm.

It is convenient to introduce the dimensionless variable
$x = k\rho$ ($k =  45.8\, x \,$MeV/$c$) and rewrite Eq.~(29)
for the differential cross section of the forward
charge-exchange deuteron breakup
in the form:
\begin{equation}
 \frac{d^2\sigma }{dt}(d + p \rightarrow (pp) + n)\Bigl|_{t=0} =
 \frac {d\sigma}{dt}(d +p \rightarrow (pp) + n) \Bigl|_{t=0}
Q(x)\,x^2 dx,
\end{equation}
where
\begin{equation}
       Q(x) = \frac{1}{2\pi^2\rho^3}\, G(k).
\end{equation}
The functions $Q(x)$ and $x^2Q(x)$ calculated with the
$s$-wave deuteron wave function according to
Hulthen expression (dotted curves) and
the square-well potential (solid curves) are shown
in Fig. 1. Their difference does not exceed a few percent
and can be taken as a measure of the theoretical uncertainty.
For comparison, the dash curves in Fig. 1 show the
functions $Q_0(x)$ and $x^2Q_0(x)$ calculated for the case
of the absent final state interaction:
\begin{equation}
  Q_0(x) = \frac{1}{2\pi^2\rho^3}\Bigl|\int \psi_d(r)\,
 \exp(-i{\bf k}{\bf r})\, d^3{\bf r}\Bigr|^2 =
  \frac{2}{\pi k^2 \rho^3}\Bigl(\int \limits_0^{\infty}
u(r) \sin kr dr \Bigr)^2.
\end{equation}
The calculation was done with
the Hulthen wave function allowing to
express the function $Q_0(x)$ in the following analytical form:
\begin{equation}
Q_0(y)= \frac{4}{\pi}\,\frac{\rho}{\rho - d} \left( \frac{1}{1 + x^2}
-  \frac{1}{\alpha^2 + x^2}\right)^2.
 \end{equation}

\section{ Contribution of the deuteron $d$-wave state}

Since the proton-neutron system in the deuteron has the
total angular momentum $J=1$ and positive parity, its total spin
$S=1$ and the orbital angular momentum $L=0$ and $2$.
The normalized wave function of the deuteron, taking into account
both the $s$- and $d$-wave contributions, has the form [5]:
\begin{equation}
   \psi^{(d)}_m = \frac{\beta}{{\sqrt 4\pi}}\, \frac{u(r)}{r}
\mid \chi_m\rangle +  \gamma\,\frac {\omega(r)}{r}\,
\sum_{\mu = 0, \pm1} C^{1\,m}_{2\,m - \mu; 1\,\mu} \,
Y_{2\,m-\mu}\bigl(\frac{{\bf r}}{r}\bigr)\, \mid \chi_{\mu} \rangle,
\end{equation}
where $C^{1\,m}_{2\,m-\mu;1\,\mu}$ are the Clebsh-Gordan coefficients.
In so doing, $\beta\approx 0.98$, $\gamma \approx 0.2$,
$$
\int\limits_0^{\infty} u^2(r)\,dr =1, \qquad
\int \limits_0^{\infty}\omega^2(r)\,dr =1, \qquad  \beta^2 + \gamma^2
=1.
 $$
The differential cross section of the forward breakup
process $d + p \rightarrow
(pp) + n $ becomes:
\begin{equation}
\frac{d^4\sigma}{dt}(d + p \rightarrow (pp) + n)\Bigl|_{t=0} =
2\,{\rm tr}\,\left(\hat{\rho}^{(p)} \sum_m\sum_{m'} \hat{M}^+_m
\hat{M}_{m'} \rho^{(d)}_{m,m'}\right)\,\frac{k^2 dk}{(2\pi)^3}
\,d\Omega_{{\bf k}}.
\end{equation}
In Eq. (39), \,$\hat{\rho}^{(p)}$ is the two-row spin density matrix of
the target proton, $\hat{\rho}^{(d)}$ is the three-row spin density
matrix of the projectile deuteron, $d\Omega_{{\bf k}}$ is the element
of the solid angle along the direction ${\bf k}$; the transition
amplitudes $ \hat{M}_m$ ($m = +1,\, 0, -1$) are the two-row matrices:
\begin{equation}
\hat{M}_m = \frac{4\pi}{k}\Bigl[\frac{\beta}{\sqrt{4\pi}}\,
g_0(k)\,\exp(i\delta_0(k))\, \hat{R}_m + \gamma\, g_2(k)
\exp(i\delta_2(k)) \sum_{\mu=0,\pm 1} C^{1\,m}_{2\,m-\mu;1
\,\mu}Y_{2\, m-\mu}(\theta, \phi)\,\hat{R}_{\mu}\Bigr],
\end{equation}
where
$$
\hat{R}_{+1} = -\,b\,\frac {\hat{\sigma}_x
+i\,\hat{\sigma}_y}{\sqrt{2}},\quad \hat{R}_0 = a\,\hat{\sigma}_z, \quad
\hat{R}_{-1} = \,b\, \frac{\hat{\sigma}_x -
i\,\hat{\sigma}_y}{\sqrt{2}};
 $$
 \begin{equation} g_0(k) =
\int\limits_0^{\infty} u(r)\,R_{k0}(r)dr, \quad g_2(k) =
\int\limits_0^{\infty} \omega(r)\,R_{k2}(r)dr ,
 \end{equation}
$\theta$ and $\phi$ are the
polar and azimuthal angles determining the direction of the proton
momentum in the two-proton rest frame (coinciding here with the
deuteron rest frame), the phases $\delta_L(k)$ of the
singlet $pp$-scattering for $L=0$ and 2
and the corresponding radial wave functions $R_{kL}(r)$
are defined in Eq.~(27).
Note that, according to Eq. (28),
$g_0^2(k) = (k^2/4\pi) G(k)$.

For unpolarized initial particles,
we get:
 $$
\frac{d^4\sigma}{dt}(d + p \rightarrow (pp) + n)\Bigl|_{t=0} =
\frac{2}{3} \Bigl\{\left[\beta^2g_0^2(k) +
 \gamma^2g_2^2(k)\right]\frac{2|b|^2 + |a|^2}{4\pi} +
 \left[\gamma^2 g^2_2(k)+\right.
 $$
 \begin{equation}
 \left.
+ 2\sqrt{2}\,\beta\,\gamma g_0(k)g_2(k)\cos\bigl(
\delta_0(k) - \delta_2(k)\bigr)\right]\bigl(|a|^2 - |b|^2
\bigr)\frac{3\cos^2\theta - 1}{8\pi}
\Bigr\}\,\frac{2}{\pi}\,dk\,d\Omega_{{\bf k}},
\end{equation}
where $\theta$ is the angle between the
momentum of one of the protons in the two-proton (deuteron)
rest frame and the deuteron beam
momentum ($\cos \theta = {\bf k}{\bf l}/k$).

We see that at $a\ne b$ the contribution of the $d$-wave state
leads to the anisotropy of the angular distribution
of the relative momenta of the two protons.
Besides, this contribution violates
the factorization of the differential cross section.
The factorization is however recovered after
the integration over the solid angle:
\begin{equation}
\frac{d^2\sigma}{dt}(d + p \rightarrow (pp) + n)\Bigl|_{t=0} =
\frac{2}{3}\bigl( 2|b|^2 + |a|^2\bigr)\,\bigl(\beta^2g_0^2(k) +
 \gamma^2g_2^2(k)\bigr)\, \frac{2}{\pi}
\,dk.
 \end{equation}
Moreover, integrating over the momentum spectrum, using the normalization
conditions
 $$
\frac{2}{\pi}\int\limits_0^{\infty} g_0^2(k)\,dk =
\int_0^{\infty} u^2(r)\,dr = 1,\qquad
\frac{2}{\pi}\int\limits_0^{\infty} g_2^2(k)\,dk =
\int_0^{\infty} \omega^2(r)\,dr = 1, \qquad \beta^2 + \gamma^2 =1,
 $$
we recover Eq.~(14), originally obtained for the $s$-wave
deuteron state:
$$
  \frac{d\sigma}{dt}(d + p \rightarrow (pp) + n) \Bigl|_{t=0} =
\frac{2}{3}\,\bigl(2|b|^2 + |a|^2\bigr) =
\frac{2}{3}\, \frac{d\sigma^{(f)}}{dt}(n + p \rightarrow p + n)
\Bigl|_{t=0}.
$$

For polarized initial particles the spectrum
of the proton momenta integrated over
the solid angle depends on the same deuteron and
proton polarization as in Eq.~(19):
$$
\frac{d^2\sigma}{dt}(d + p \rightarrow (pp) + n)\Bigl|_{t=0} =
2\,\Bigl\{\frac {2|b|^2 + |a|^2}{3}\, \bigl(\beta^2 g_0^2(k) +
\gamma^2 g_2^2(k)\bigr)-
$$
$$
-\Bigl[|b|^2 (P^{(d)}_{\parallel}P^{(p)}_{\parallel}+
{\mathrm Re}(ba^*)\,({\bf P^{(d)}_{\perp} P^{(p)}_{\perp}})\Bigr]
\bigl(\beta^2g_0^2(k) - \frac{1}{2}\, \gamma^2g_2^2(k)\bigr)+
$$
\begin{equation}
+ \Bigl[\bigl(|b|^2 - |a|^2\bigr)\,t^{(d)}_{2\,0} + {\mathrm Im}(ba^*)
({\bf t_{\perp}^{(d)} [P^{(p)}l]})\Bigr]\bigl(\beta^2g_0^2(k) +
\frac{1}{10} \gamma^2 g_2^2(k)\bigr)\Bigr\}\,\frac{2}{\pi}\,dk.
\end{equation}
Integrating further over the
spectrum of proton momenta, we recover
Eq.~(19) with the deuteron polarization parameters:
 $P^{(d)}_{\parallel}$, ${\bf P}^{(d)}_{\perp}$, $t^{(d)}_{2\,0}$,
${\bf t}_{\perp}^{(d)}$ substituted by the polarization parameters of the
neutron-proton triplet state in the deuteron:
 $P^{(np)}_{\parallel}$, ${\bf P}^{(np)}_{\perp}$, $t^{(np)}_{2\,0}$,
 ${\bf t}_{\perp}^{(np)}$.
The elements of the spin density matrix of this state
are connected with the elements of the spin density matrix of the
deuteron by the following simple relations:
\begin{equation}
\rho^{(np)}_{\mu,\mu'} = \beta^2 \rho^{(d)}_{\mu,\mu'} + \gamma^2
 \sum_M C^{1\,M + \mu}_{2\,M;1\,\mu}\,C^{1\,M + \mu'}_{2\,m;1\,\mu'}
\rho^{(d)}_{M + \mu,\, M + \mu'}.
\end{equation}
Accordingly, the polarization parameters are connected by the
relations:
\begin{equation}
{\bf P}^{(np)} = \bigl(\beta^2 - \frac{1}{2}\gamma^2\bigr)\,
{\bf P}^{(d)};~~~
t^{(np)}_{2\,0}=\bigl(\beta^2 +
\frac{1}{10}\gamma^2\bigr)\,t^{(d)}_{2\,0};~~~
{\bf t}_{\perp}^{(np)} =
\bigl(\beta^2 + \frac{1}{10}\gamma^2\bigr)\,{\bf t}_{\perp}^{(d)}.
\end{equation}

It should be noted that we have not considered the Glauber
corrections due to the rescattering.
The estimates with the Hulthen wave function of the
deuteron show that  at small relative proton momenta
($k < 50$ MeV/$c$) these corrections do not exceed several percent.
However, they can be essential ($\sim 20\,\%$) and should be taken
into account for an upper part of the momentum spectrum ($ k > 100$
MeV/$c$).
\section {Summary}
~~~
1. In the impulse approach, the relation between the
 differential cross section of the charge-exchange breakup of a fast
 deuteron with the production of two protons and the differential
cross section of the neutron-proton charge transfer reaction
has been established.

2. The role of the effects of the proton identity
(Fermi-statistics) and the proton interaction in the final state
has been clarified. Particularly, the identity effect forces the
two-proton system to be produced in the forward charge-exchange
deuteron breakup process $ d + p \rightarrow (pp) + n$
only in the singlet state.
As a result, the differential cross section
of this process is determined only by the spin-dependent part of the
amplitude of the reaction $n + p \rightarrow p + n $ at zero
angle.

3. The study of the charge-exchange deuteron
breakup with a beam of polarized (aligned) deuterons
on the polarized hydrogen
target allows one to separate two spin-dependent terms in the
amplitude of the charge transfer reaction $n + p \rightarrow p + n$
and to determine the phase difference between them.

4. The spectrum of the proton relative momenta in the
forward charge-exchange deuteron breakup process
$d + p \rightarrow (pp) + n$ has been calculated with the
account of the Coulomb and strong interactions of the protons in
the final state. The accuracy of the calculation of the soft part
of the spectrum ($k < 50$ MeV/$c$) is estimated
on the level of several percent.

5. The influence of  the $d$-wave state of the deuteron
on both the
proton momentum spectrum and the polarization effects
has been analyzed.\\

This work was supported by GA Czech Republic, Grant. No. 292/01/0779,
by Russian Foundation for Basic Research, Grant No. 01-02-16230,
and within the Agreements IN2P3-ASCR No. 00-16 and IN2P3-Dubna No.
00-46.

\begin{figure}
\begin{center}\mbox{\epsfig{file=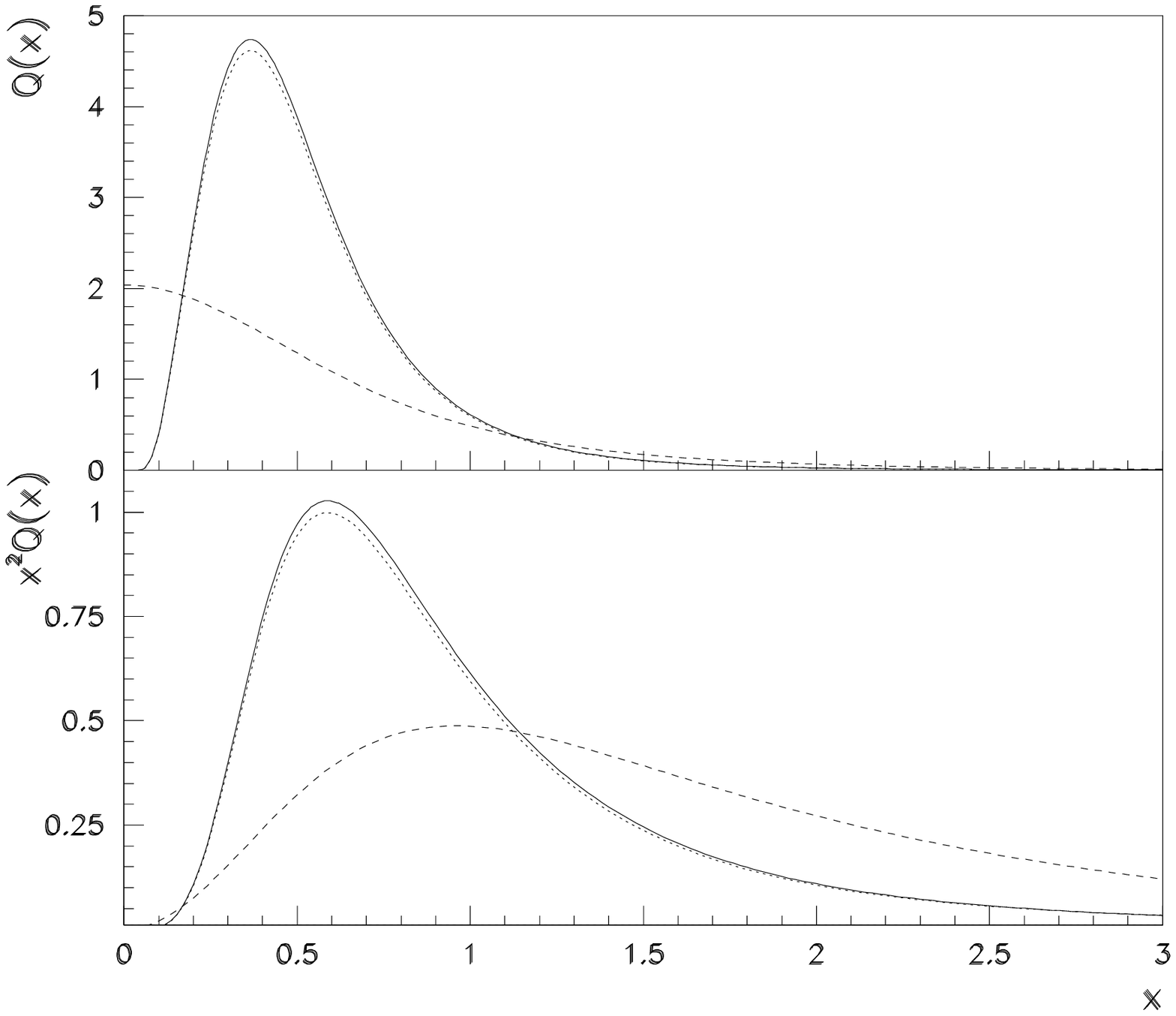,height=12cm}}\end{center}
\caption{The spectra of the proton relative momenta in the
forward charge-exchange deuteron breakup
($k =  45.8\, x \,$MeV/$c$, see Eq.~(34))
calculated with the
$s$-wave deuteron wave function according to
Hulthen expression (dotted curves) and
the square-well potential (solid curves).
The dash curves correspond to the case
of the absent final state interaction.}
\label{fig1}
\end{figure}

\end{document}